\documentclass[]{pasj00}
\Received{2009 March 31}
\Accepted{2009 June 16}
\Published{$\langle$publication date$\rangle$}
\SetRunningHead{T. Yuasa et al.}{The origin of an extended X-ray emission of 47 Tuc}

\newcommand{\countss}{{\rm counts}\ {\rm s}^{-1}}
\newcommand{\ergs}{{\rm erg}\ {\rm s}^{-1}}
\newcommand{\ergcms}{{\rm erg}\ {\rm cm}^{-2}\ {\rm s}^{-1}}
\newcommand{\ergcmssr}{{\rm erg}\ {\rm cm}^{-2}\ {\rm s}^{-1}\ {\rm sr}^{-1}}
\usepackage{times}

\begin{document}

\title{The origin of an extended X-ray emission apparently associated with the globular cluster 47 Tucanae}
\author{
Takayuki \textsc{Yuasa},\altaffilmark{1}
Kazuhiro \textsc{Nakazawa},\altaffilmark{1}
and Kazuo \textsc{Makishima}\altaffilmark{1,2}
\thanks{Last update: \today}}
\altaffiltext{1}{Department of Physics, School of Science, University of Tokyo, 7-3-1 Hongo, Bunkyo-ku, Tokyo 113-0033}
\altaffiltext{2}{Cosmic Radiation Laboratory, The Institute of Physical and Chemical Research (RIKEN), 2-1 Hirosawa, Wako, Saitama 351-0198}
\email{yuasa@juno.phys.s.u-tokyo.ac.jp}

\maketitle

\begin{abstract}
Using the Suzaku X-ray Imaging Spectrometer, we performed a 130 ks observation of an extended X-ray emission, which was shown by ROSAT and Chandra observations to apparently associate with the globular cluster 47 Tucanae. The obtained $0.5-6$~keV spectrum was successfully fitted with a redshifted thin thermal plasma emission model whose temperature and redshift are $2.2^{+0.2}_{-0.3}~$keV (at the rest frame) and $0.34\pm0.02$, respectively. 
Derived parameters, including the temperature, redshift, and luminosity, indicate that the extended X-ray source is a background cluster of galaxies, and its projected location falls, by chance, on the direction of the proper motion of 47 Tucanae.
\end{abstract}

\section{Introduction}
Many globular clusters in the Galaxy move through the Galactic halo with a typical velocity of $\sim 200$ km s$^{-1}$, which exceeds the sound velocity (a few tens of km~s$^{-1}$ to 150~km~s$^{-1}$) specified by roughly estimated halo plasma temperatures ($T\sim0.7-1.4\times10^5$~K by \cite{sdb81}; $T\sim1.5-1.6\times10^6$~K by \cite{pie98}). Then, a bow shock is expected to form between the halo plasma and any gas (intra-cluster gas) in a moving globular cluster \citep{rs71}.
Since the temperature of the post-shock plasma should then become $\sim10^6$ K (e.g., \cite{oka07}), we expect to detect diffuse X-ray emission with a shape that traces the bow shock. 
Several X-ray satellites have been observing globular clusters in search for such bow-shock-heated X-ray emitting plasmas. 
Indeed, using the Einstein satellite, \citet{har82} first detected such diffuse emissions around 47 Tucanae (hereafter 47 Tuc), $\rm{\omega}$~Centauri, and M22, although a part of the emissions was resolved into point sources by later observations (\cite{ka87, kg95, gen03}). Subsequently, \citet{kg95} newly reported diffuse emissions in 47 Tuc,  followed by, possible detections of such a diffuse emission in several globular clusters; e.g., \citet{hop00} in NGC 6779, and \citet{oka07} in 47 Tuc, NGC 6752, and M5.

Among the extended emissions so far detected, those in the globular clusters 47 Tuc and NGC 6752 are of particular interest.
They spatially coincide with the directions of projected proper motions of these globular clusters \citep{kg95,oka07}, and were hence considered to have physical relationships with the globular clusters.
According to \citet{oka05} and \citet{oka07}, the 0.5-4.5~keV Chandra spectra from these extended sources are so hard that they require power-law models with photon indices of $\Gamma=2.1\pm0.3$ (47 Tuc) and $\Gamma=2.0\pm0.2$ (NGC 6752), or a thermal plasma emission model with temperatures of $kT=3.7^{+2.7}_{-1.3}$~keV (47 Tuc) and $kT=2.9^{+1.0}_{-0.7}$~keV (NGC 6752) which largely exceed values expected from the bow shock heating ($\sim10^6$~K).
While the two X-ray sources have no optical identifications \citep{kg95,oka07}, \citet{oka07} reported that both have possible radio counterparts in the 843~MHz Sydney University Molongo Sky Survey (SUMSS; \cite{boc99}). 

From these properties, the extended emissions apparently associated with 47 Tuc and NGC 6752 were considered to arise via inverse Compton scattering \citep{kg95} or non-thermal bremsstrahlung \citep{oka07} of high energy ($E\sim20-100$~keV) electrons that are stochastically accelerated in the bow shock. 
The interpretation is attractive because the shock is expected to be a moderate one with a Mach value of $\sim10$, and the condition is much different from those in the more typical acceleration sites such as supernova remnants and jets of active galactic nuclei.

As an alternative explanation to those extended X-ray sources, \citet{kg95} and \citet{oka07} also considered a chance coincidence with a background cluster of galaxies that is not related to the globular clusters. This alternative must be kept in mind, even though its possibility was estimated low ($<1\%$ by \cite{kg95} and \cite{oka07}) in 47 Tuc.

In the previous spectral analysis of the Chandra data from the extended emission in 47 Tuc and NGC 6752, \citet{oka07} were unable to distinguish a power-law model from a thermal emission model because of rather large statistical errors.
In the present paper, we utilize the larger effective area and lower background level of Suzaku, to perform detailed spectral analysis of the extended emission of 47 Tuc. Based on the model fitting result, we conclude that the emission is from a background galaxy cluster with a redshift of 0.3, and the spatial coincidence between the extended emission and the globular cluster is accidental.

Throughout the paper, cosmological parameters of $\Omega_{\rm{M}}=0.28$ and $H_0=70$~km~s$^{-1}$~Mpc$^{-1}$ are used in calculations.

\section{Observation and Data Reduction}\label{sec:observation_datareduction}
The globular cluster 47 Tucanae was observed with Suzaku \citep{mit07} on 2007 June 10--12 (observation ID 502048010). Since the target of this observation is the extended emission (EE) at the north east region of 47 Tuc, we set the nominal pointing of the satellite at $(\timeform{00h24m50s}, \timeform{-71D59'46''})$, which is $\sim\timeform{6'}$ off the cluster center.

In the present study, we focus on the data taken with the X-ray Imaging Spectrometer (XIS; \cite{koy07}), which comprises four X-ray charge coupled device (CCD) sensors each placed on the focal plane of the X-ray Telescope (XRT; \cite{ser07}). The four pairs of XIS and XRT are co-aligned together, and have the same field of view (FOV) of $\timeform{17.8'}\times\timeform{17.8'}$. Since one of three front-side illuminated (FI) CCD chips, XIS2, was not operational since 2006 November, we utilized the data from the remaining two FI sensors (XIS0 and 3) and a back-side illuminated (BI) one (XIS1). In the present observation, the XIS was operated in the normal mode without any window or burst option, but incorporating the spaced-row charge injection method \citep{nak08} to restore the energy resolution of the CCDs.

After removing periods of the Earth elevation angle less than $5^{\circ}$ (ELV$<5^{\circ}$), the day Earth elevation angle less than  $20^{\circ}$ 
(DYE$\_$ELV$<20^{\circ}$), and the South Atlantic Anomaly, we achieved a net exposure of 132 ks. Flickering pixels were removed from the data by using \verb|cleansis| version 1.7. Then, cleaned event files were generated employing the same event extraction criteria as in the Suzaku pipe line processing (version 2).

The present data reduction and analysis were performed using HEADAS package version 6.4.1 and \verb|XSPEC| version 11.3.2. In spectral fitting, redistribution matrix files and ancillary response files (ARFs) for the XIS/XRT were generated using \verb|xisrmfgen| version 2007-05-14 and \verb|xissimarfgen| \citep{ish07} version 2008-04-05, respectively, with the calibration files which are provided by the calibration database (CALDB) version 2008-04-01. In the spectral fitting described below, we ignored data in the $1.8-1.9$~keV band so as to avoid calibration uncertainties around the Si-K edge.

Events with energies above 10~keV, taken with the Hard X-ray Detector (HXD; \cite{tak07}), were not utilized in the present analysis. This is because the HXD lacks imaging capability, and hence we cannot exclude contamination by X-rays from a number of point sources associated with 47 Tuc (eg. \cite{ver98,gri01,hei05}).

Since the spatial resolution of the XIS/XRT is $\sim \timeform{2'}$, we cannot exclude, using the XIS data alone, X-ray point sources that overlap with the EE. To determine their spectral shapes and fluxes, we also utilized archived Chandra ACIS data of 47 Tuc acquired in 2000 March for a total exposure of about 70~ks (obsid=953 and 955). We used \verb|CIAO| (Chandra Interactive Analysis of Observations) version 4.0.2 and CALDB version 3.4.5 to extract point source spectra. Like in the Suzaku data analysis, we also used \verb|XSPEC| when performing model fitting to the ACIS spectra.

\section{Image Analysis}\label{sec:imgana}
\subsection{Soft and hard band images}\label{sec:bandimages}
\begin{figure*}
  \begin{center}
    \FigureFile(170mm,70mm){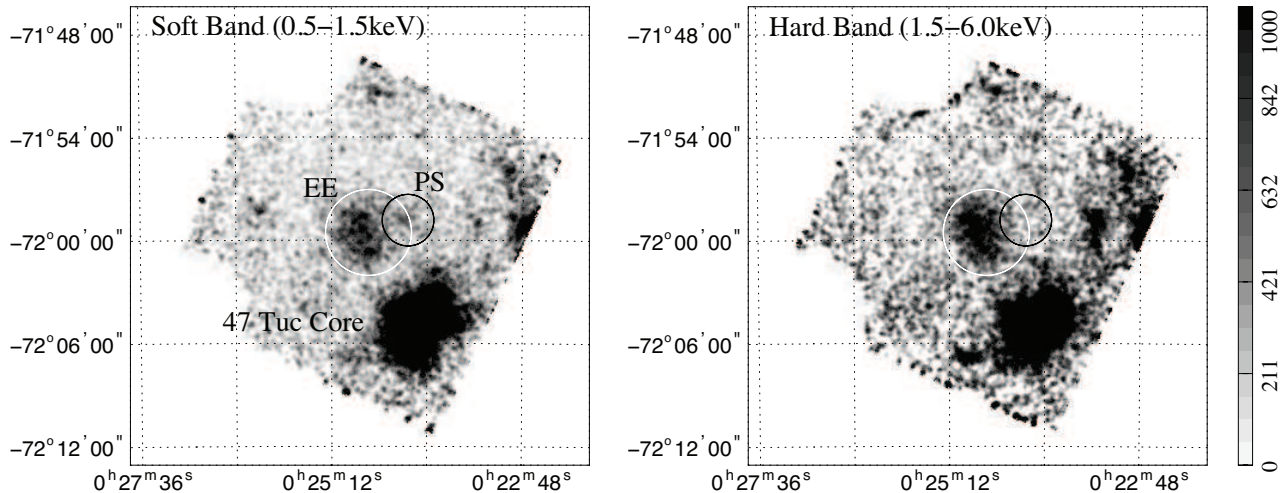}
  \end{center}
  \caption{Soft ($0.5-1.5$~keV; left) and hard ($1.5-6.0$~keV; right) band images of 47 Tuc taken with XIS FI (XIS0 plus XIS3), shown after removing the two corner regions irradiated with the calibration source. The images are scaled in units of $4\times 10^{-5} \countss~{\rm pixel}^{-1}$. The non X-ray background is subtracted using the night Earth image (see text), followed by vignetting and exposure correction, although the diffuse X-ray background is included. The white circle with a radius of $\timeform{150''}$ is the event extracting region for the extended emission (EE), and the black small one shows a soft point source (PS) which is excluded from the spectral analysis.}
  \label{fig:xis_image}
\end{figure*}

In Fgure \ref{fig:xis_image}, we present images obtained with the FI cameras (XIS0 and 3) in the soft ($0.5-1.5$~keV) and hard ($1.5-6.0$~keV) bands, after subtracting the non X-ray background (NXB) and correcting for vignetting and exposure.
We estimated the NXB of the XIS using dark (night) Earth data taken within $\pm150$ days of our observation of 47 Tuc. The night Earth data were summed up, with weights according to geomagnetic cut-off rigidity which the spacecraft experienced at the data acquisition. This was performed by \verb|xisnxbgen| \citep{taw08}. Then, we created NXB images in the soft and hard bands, and subtracted them from the raw images. After subtracting the NXB, we smoothed each image with a two-dimensional Gaussian of $\sigma=\timeform{6''}$. The diffuse X-ray backgrounds, namely the cosmic X-ray background (CXB) and Galactic diffuse emission, are still included in the images.

In figure \ref{fig:xis_image}, we see several point sources, and the very bright 47 Tuc core region which consists of numerous X-ray point sources (e.g., \cite{hei05}). At the center of the two images, we also observe a clear concentration of X-ray events as \citet{kg95} and \citet{oka07} reported. Thus, we reconfirm the EE phenomenon with the Suzaku data.
To extract photons from the EE region, we define a circular region (white circle in figure \ref{fig:xis_image}) with a radius of $\timeform{150''}$, centered on $(\timeform{00h24m44.2s}, \timeform{-71D59'33.5''})$ where \citet{oka07} found the maximal surface brightness.

As indicated with a black solid circle and a label ``PS" in figure \ref{fig:xis_image}, a faint point source is  recognized at the north west rim of the event extracting region. Although the EE is still apparent in the hard band image, the point source is no longer visible therein. 
At a consistent position, we find a point source also in the ACIS image. Therefore, we consider that the XIS source is not a brightness enhancement associated with the EE, and hereafter exclude it using a circular region with a radius of $\timeform{1'}$. This region is expected to enclose 50\% of X-ray photons from the point source, while the remaining half will fall out of the region; a half of those photons (25\% of the total flux from the source) are in turn estimated to fall inside the event extracting region around the EE, and contaminate the EE spectrum. This effect is considered later in section \ref{sec:specana}.

\subsection{Radial profile of the extended emission}
\begin{figure}
  \begin{center}
    \FigureFile(80mm,80mm){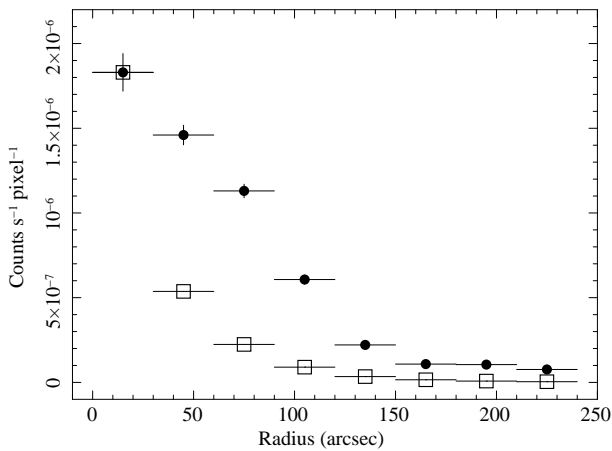}
  \end{center}
  \caption{Azimuthly averaged radial profiles of the EE (filled circles) shown in units of counts s$^{-1}$ pixel$^{-1}$, after subtracting the CXB and Galactic diffuse components (see text). The PSF of XIS FI (open squares) is also plotted, being normalized to have the same maximum value as that of the EE at the innermost annulus.}
  \label{fig:radpro}
\end{figure}

In the previous studies by \citet{kg95} and \citet{oka07}, the EE was concluded to be extended over an angular radius of $\sim\timeform{2'}$ (2.7~pc assuming a 4.6~kpc distance to 47 Tuc). To examine the spatial extent of the emission, we calculated its azimuthly averaged radial profile using the NXB-subtracted and vignetting-corrected $0.5-6.0$~keV XIS FI image. This was done utilizing a series of annular extracting regions, each with $\timeform{30''}$ width, which are concentric with the original event extraction region. The result is shown in figure \ref{fig:radpro} after subtracting the CXB and Galactic diffuse background rate of $9.4\times10^{-7}~\countss~{\rm pixel}^{-1}$, which we estimated using another region of the CCDs with no evident point sources.

In the same figure, we also plot a radial profile of the point spread function (PSF) of the XIS/XRT, calculated at the center of the EE, and averaged over XIS0 and XIS3. Thus, the EE is clearly more extended than the PSF even though the latter is much broader than those of ROSAT and Chandra.

\section{Spectral Analysis}\label{sec:specana}
\subsection{Extraction of spectra}
We extracted XIS FI and BI spectra of the EE using the event extracting region shown in figure \ref{fig:xis_image} (the white circular region but excluding the black circle).
The $0.5-6$~keV band count rates measured with XIS FI and BI are $28.3\pm0.05\times10^{-3}~\countss$ and $22.7\pm0.04\times10^{-3}~\countss$, respectively, with $1\sigma$ statistical errors.

In addition to the EE which is the subject of the present analysis, the spectra also contain events from the NXB, the Galactic and extragalactic X-ray backgrounds (altogether, diffuse X-ray background), and X-ray events from several contaminating X-ray point sources that cannot be resolved with the XIS/XRT spatial resolution.
We assume the Galactic diffuse emission to have a uniform brightness across the XIS FOV.
Although its hard component \citep{wor82,koy86,ebi01,rev06, kri07} has a strong concentration toward the Galactic plane (with a scale height of $\timeform{1.5D}-\timeform{3D}$~; e.g., \cite{rev06}, \cite{kri07}), and hence a steep brightness gradient, it can be neglected at this high Galactic latitude of $\sim\timeform{45D}$ of 47 Tuc.

\subsection{Background Spectra}
In the following subsections, we estimate the spectral shapes and fluxes of these individual background components, and create XIS spectral data for each component. By summing all these components, we construct background spectra, and then subtract them from the raw XIS spectra of the EE. 

\subsubsection{The non X-ray Background}\label{sec:nxb}
We derived the NXB spectrum from the same stacked night-Earth data as described in section \ref{sec:bandimages}. Since this component depends on the CCD location, we examined spectral differences among several circular extracting regions on the night-Earth image, with the radius ranging from $\timeform{150''}$ to $\timeform{300''}$. Each region is concentric with the EE extracting region (white circle in figure \ref{fig:xis_image}).
The derived $0.5-10$~keV NXB spectra were consistent with one another within $\sim3$\%.
Therefore, to minimize the statistical errors of the estimated NXB, we adopted the largest extracting region ($\timeform{300''}$ radius) for both XIS FI and BI.

The constructed NXB spectrum is shown in figure \ref{fig:bgdsummary} in green.
The count rate has been scaled to the ratio ($\sim4.8$) of the NXB and signal extracting areas.
The $0.5-6$~keV band count rates with $1\sigma$ statistical errors are $4.4\pm0.1\times10^{-3}~\countss$ (XIS FI) and $5.3\pm0.1\times10^{-3}~\countss$ (XIS BI).

\subsubsection{The diffuse X-ray background}
The diffuse X-ray background consists of two components; the Galactic and the extragalactic emissions. The former component is thought to originate from the Galactic halo and the Local Hot Bubble (eg. \cite{cox87}), and expected to appear at energies below $\sim2$~keV with its surface brightness depending considerably on the sky direction. The latter, the extragalactic component, has been understood as a superposition of numerous extragalactic active Galactic nuclei. The spectrum is known to be expressed by a power-law model with a photon index of $\Gamma=1.4$ (e.g. \cite{par99,lum02,kus02}) at least over the $2-10$~keV band.

\begin{figure}
  \begin{center}
    \FigureFile(65mm,65mm){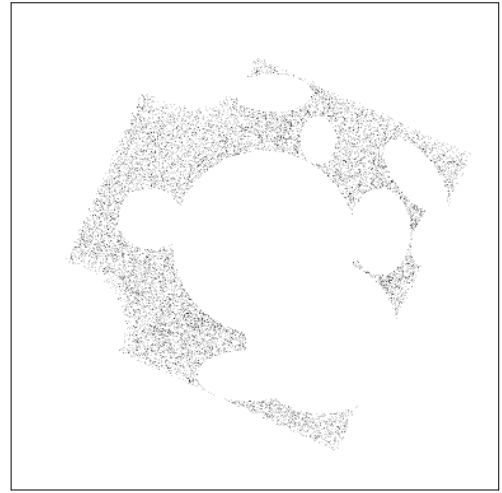}
  \end{center}
  \caption{The XIS FI image after filtering out point sources, the 47 Tuc core region, and the EE. The image is not corrected for the vignetting or exposure. The events plotted in the image were used in the modeling of the diffuse X-ray background in the field of 47 Tuc.}
  \label{fig:mask}
\end{figure}

\begin{figure}
  \begin{center}
    \FigureFile(80mm,80mm){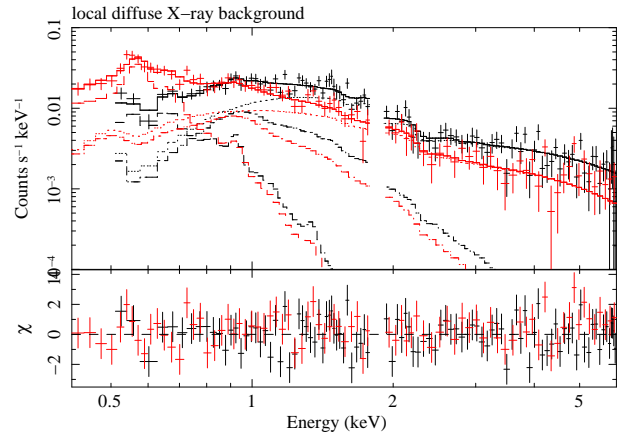}
  \end{center}
  \caption{The NXB-subtracted diffuse X-ray background spectra of XIS FI (black) and BI (red), extracted from the image in figure \ref{fig:mask}. The solid lines represent the best fit model, while their components are individually plotted in dashed (thermal), dot-dashed (thermal with absorption), and dotted (power law with absorption) lines.}
  \label{fig:diffusebgd}
\end{figure}

In order to determine the local diffuse X-ray background in the present XIS FOV, we extracted another set of XIS FI and BI spectra from the same observation data set of 47 Tuc, but applying a mask which excludes point sources, the core region of 47 Tuc, and the EE itself. The masked image of XIS FI is shown in figure \ref{fig:mask}, and the NXB-subtracted (as described in section \ref{sec:nxb}) spectra of the diffuse X-ray background are plotted in figure \ref{fig:diffusebgd}.
We fitted these spectra jointly with a model which consists of three diffuse X-ray background components;
a thermal emission from the Local Hot Bubble plasma (\verb|mekal| model in \verb|XSPEC|; \cite{lie95});
a thermal emission from the Galactic halo plasma (\verb|mekal|); and
a power-law model with a fixed photon index of 1.4 to account for the extragalactic component (\verb|powerlaw|).
The latter two components were assumed to suffer the line-of-sight Galactic absorption, with the absorbing column density fixed at $5\times10^{20}$ atoms cm$^{-2}$ \citep{dl90} which is a typical value toward the present field. The photoelectric absorption coefficient by \citet{mm83}, \verb|wabs| model in XSPEC, was employed.
We assumed that the diffuse background has a uniform surface brightness over the XIS FOV, and utilized an ARF which was calculated using \verb|xissimarfgen| with the \verb|UNIFORM| option and \verb|r_max|$=\timeform{20'}$. We left free the temperatures, metal abundances, and surface brightnesses of the two thermal models, as well as the photon index and surface brightness of the power-law model.
The model gave an acceptable fit with $\chi^2_{\nu}=1.10~(\nu=175)$; the best fit parameters are listed in table \ref{tab:localcxbmodel4}.

As to the power-law component, the best fit model gave the $2-10$~keV flux of $4.4\times10^{-8}~\ergcmssr$, which is $\sim20$\% lower than the previously reported values; $5.4\times10^{-8}~\ergcmssr$ by \citet{lum02}, and $5.7\times10^{-8}~\ergcmssr$ by \citet{kus02}.
The deviation can reasonably be explained by the spatial fluctuation of the extragalactic emission which can vary by about 20\% \citep{kus02} depending on XIS pointings.

We might directly subtract the diffuse background spectrum of figure \ref{fig:diffusebgd} from that of the EE region. However, this introduces some systematic bias because the energy-dependent vignetting effect of the XRT (figure 11 of \cite{ser07}) will cause not only the observed background brightness but also its spectral shape to differ between the two regions;
 in the present case,  the two ARFs, one for the masked region (figure \ref{fig:mask}) while the other for the EE, differ by $10-20$\% (due to energy dependent vignetting effect) in the $2-6$~keV range if we normalize them at 2~keV.
Hence, to avoid this problem, we simulated the expected contribution of the diffuse background to the EE extracting region using the best fit model explained above and the corresponding ARF.
In producing the fake spectra, we assumed a sufficiently long exposure ($10^{7}$~s), to suppress statistical errors. This is allowed because the background components are understood from previous observations.
The $0.5-6$~keV band count rates of the faked spectra are $8.3\times10^{-3}~\countss$ (XIS FI) and $5.9\times10^{-3}~\countss$ (XIS BI).

\begin{table}
\caption{The best fit model parameters for the diffuse background spectra.}\label{tab:localcxbmodel4}
\begin{center}
\begin{tabular}{ccc}
\hline
Model&Parameter&Value\\ \hline
Thermal 1&$kT$& $0.17\pm^{+0.01}_{-0.02}~{\rm keV}$\\
&$Z$\footnotemark[$*$]& $0.1\pm^{+0.9}_{-0.03}$\\
& $\Sigma$\footnotemark[$\dagger$] & $1.07\pm^{+0.51}_{-0.98}$\\%+47.8% -92.0%
Absorption &$N_{\rm H}$\footnotemark[$\ddagger$]& $5~({\rm fixed})$\\
Thermal 2&$kT$& $0.78^{+0.30}_{-0.39}~{\rm keV}$\\
&$Z$\footnotemark[$*$]& $0.03^{+0.04}_{-0.03}$\\
& $\Sigma$\footnotemark[$\dagger$] & $0.75^{+2.28}_{-0.35}$\\%%+300.25% -15.3%
Power law&$\Gamma$& $1.4~({\rm fixed})$\\
& $\Sigma$\footnotemark[$\dagger$] & $0.38^{+0.03}_{-0.03}$\\%+7.84% -7.84%
\hline
& $\chi^2_\nu$ & 1.10 (175)\\
\hline
\multicolumn{3}{@{}l@{}}{\hbox to 0pt{\parbox{70mm}{\footnotesize
\vspace{0.2cm}
\footnotemark[$*$]Abundance in terms of the solar value \citep{ag89}.\\
\footnotemark[$\dagger$]The $0.5-6$~keV band model surface brightness in units of $10^{-8}~\ergcmssr$. Absorption is not corrected.\\
\footnotemark[$\ddagger$]Line-of-sight hydrogen column density in units of $10^{20}~{\rm cm}^{-2}$.
}\hss}}
\end{tabular}
\end{center}
\end{table}

\subsubsection{Contamination from point sources}
In the previous study using Chandra \citep{oka07}, six faint X-ray point sources were found within $\timeform{2.5'}$ of the EE region. In table \ref{tab:pointsource}, we list their positions. Although they were successfully removed in the Chandra case, we cannot do so from the present XIS data except for the brightest one described in section \ref{sec:imgana}, because of the broader PSF of the XRT than that of Chandra.
Therefore, we must model and subtract their contributions, like the diffuse background. As explained below, we estimate the contribution from the soft point source (Source 1; figure \ref{fig:xis_image}) using the Suzaku XIS data themselves, and those of the remaining five point sources (Source 2--5) using the Chandra ACIS data assuming that they are not variable.

\begin{table}
\caption{The coordinates of the contaminating point sources.}\label{tab:pointsource}
\begin{center}
\begin{tabular}{ccc}
\hline
Source \#&Coordinate\\ \hline
1&$(\timeform{00h24m14.51s}, \timeform{-71D58'50.4''})$\\
2&$(\timeform{00h24m38.71s}, \timeform{-72D00'46.3''})$\\
3&$(\timeform{00h24m34.83s}, \timeform{-72D00'40.2''})$\\
4&$(\timeform{00h24m30.26s}, \timeform{-72D00'33.8''})$\\
5&$(\timeform{00h25m00.70s}, \timeform{-71D59'59.9''})$\\
6&$(\timeform{00h24m42.63s}, \timeform{-71D59'22.3''})$\\
\hline
\multicolumn{2}{@{}l@{}}{\hbox to 0pt{\parbox{70mm}{\footnotesize
\vspace{0.2cm}

}\hss}}
\end{tabular}
\end{center}
\end{table}

\begin{figure}
  \begin{center}
    \FigureFile(80mm,80mm){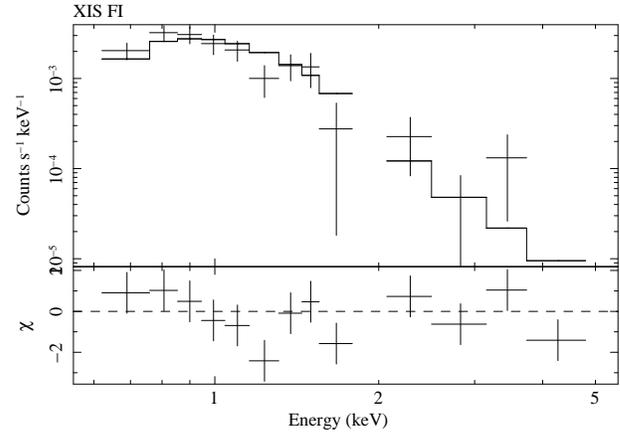}
  \end{center}
  \caption{The XIS FI spectrum (black crosses) and the best fit power-law model (solid line) of the soft point source at the north west of the EE. The NXB and diffuse X-ray background are subtracted. Data from XIS BI are excluded from the plot for clarity, although they were incorporated in the fitting.}
  \label{fig:xissoftps}
\end{figure}

Figure \ref{fig:xissoftps} shows XIS FI spectrum of Source 1, extracted from the black circle (figure \ref{fig:xis_image}), shown after subtracting the NXB and the diffuse X-ray background. The FI and BI spectra were fitted with an absorbed single power-law model in the $0.5-5$~keV band. As listed in table \ref{tab:ps}, this gave an acceptable fit with $\chi^2_\nu=1.10~(\nu=30)$.
The summed spectrum of the remaining 5 point sources was extracted from the ACIS data (section \ref{sec:observation_datareduction}), using circular regions each $\timeform{5''}$ in radius.
The NXB was extracted from another region of the same ACIS CCD with no evident point sources. Then, we fitted the summed spectrum with a single power-law model in the $0.8-6$~keV band. The best-fit ($\chi^2_\nu=2.54$ and $\nu=4$) model gave a null hypothesis probability of $0.041$, and the parameters as listed in table \ref{tab:ps}.

To obtain a summed contribution of all the point sources to the EE, we then faked the summed spectrum of the 5 point sources and the soft source separately, by applying appropriate ARFs to the best fit models described above. In calculating the ARF for the 5 point sources, we took an average of individual ARFs weighted by their $0.5-5$~keV ACIS count rates. The ARF for Source 1 was calculated referring to the XIS/XRT effective area for X-ray photons which leak into the EE event extracting region; the source position was set to be that of the soft source (the first row in table \ref{tab:pointsource}), whilst it is located outside the EE region (white circle in figure \ref{fig:xis_image})\footnote{A ratio of the number of photons which leak into the EE region to that of photons falling inside the Source 1 region (black circle in figure \ref{fig:xis_image}) is 23\% in the 0.5-5~keV band, which is close to with the rough estimation ($\sim25$\%) in section \ref{sec:bandimages}. }.
Based on the faked spectrum, the implied $0.5-6$~keV band count rates are $2.9\times10^{-3}~\countss$ (XIS FI) and $2.3\times10^{-3}~\countss$ (XIS BI). 

\begin{table}
\caption{The best fit model parameters for the contaminating soft point source and five faint sources.}\label{tab:ps}
\begin{center}
\begin{tabular}{cccc}
\hline
Model & Parameter & Source 1 & Source $2-5$\\ \hline
Absorption &$N_{\rm H}$\footnotemark[$*$]& $2.2^{+2.3}_{-1.3}$ & 0 (fixed)\\
Power law&$\Gamma$& $5.1^{+2.3}_{-1.1}$ & $1.7\pm0.2$\\
& ${\rm flux}$\footnotemark[$\dagger$] & $3.9^{+6.1}_{-1.6}$ & $3.2\pm0.3$\\%+158.3% -42.4%
\hline
& $\chi^2_\nu$ & 1.10 (30) & 2.54 (4)\\
\hline
\multicolumn{4}{@{}l@{}}{\hbox to 0pt{\parbox{70mm}{\footnotesize
\vspace{0.2cm}
\footnotemark[$*$]Hydrogen column density in units of $10^{21}~{\rm cm}^{^2}$.\\
\footnotemark[$\dagger$]The $0.5-6$~keV band model flux in units of $10^{-14}~\ergcms$. Not corrected for the absorption.
}\hss}}
\end{tabular}
\end{center}
\end{table}

\begin{figure}
  \begin{center}
    \FigureFile(80mm,80mm){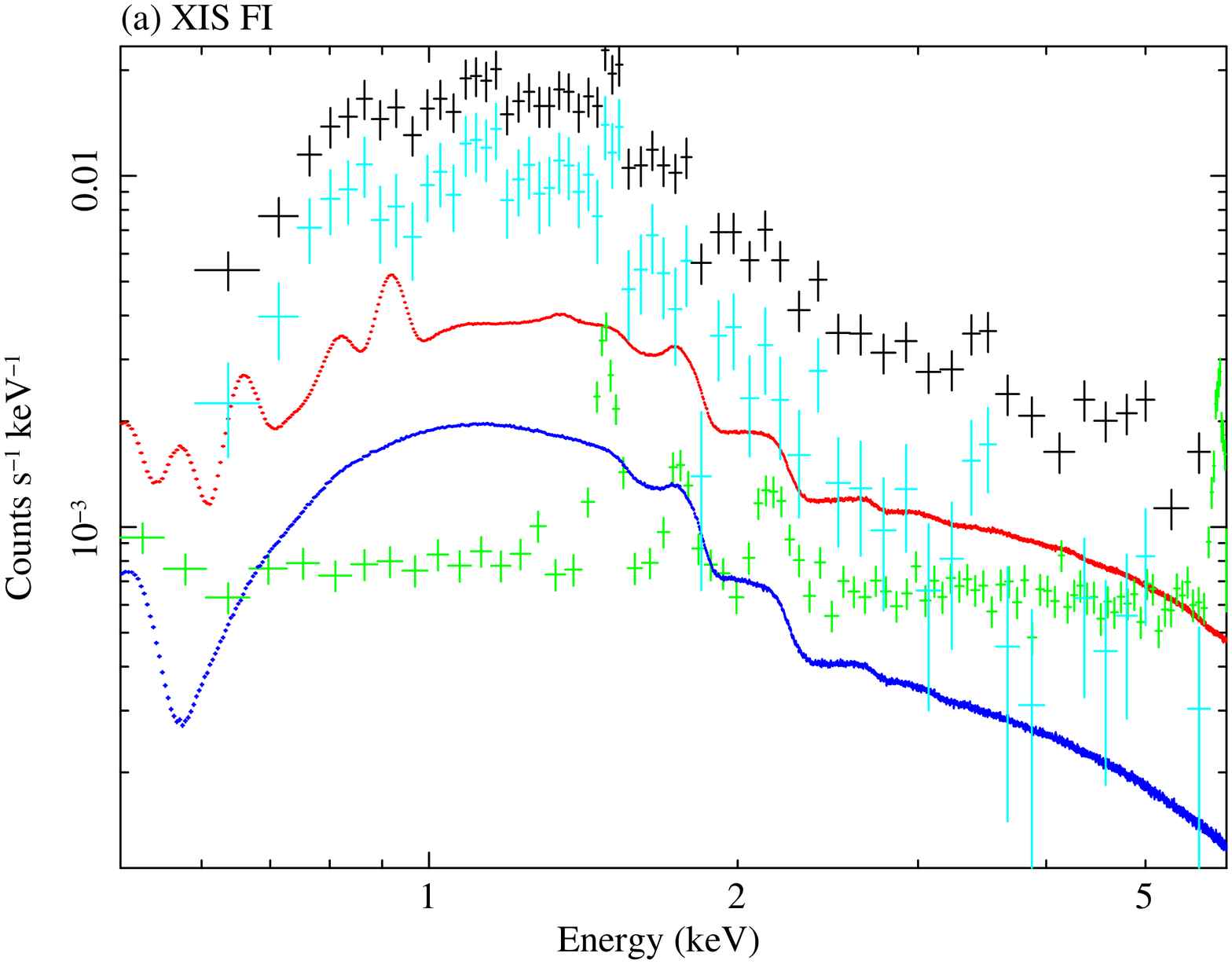}
    \FigureFile(80mm,80mm){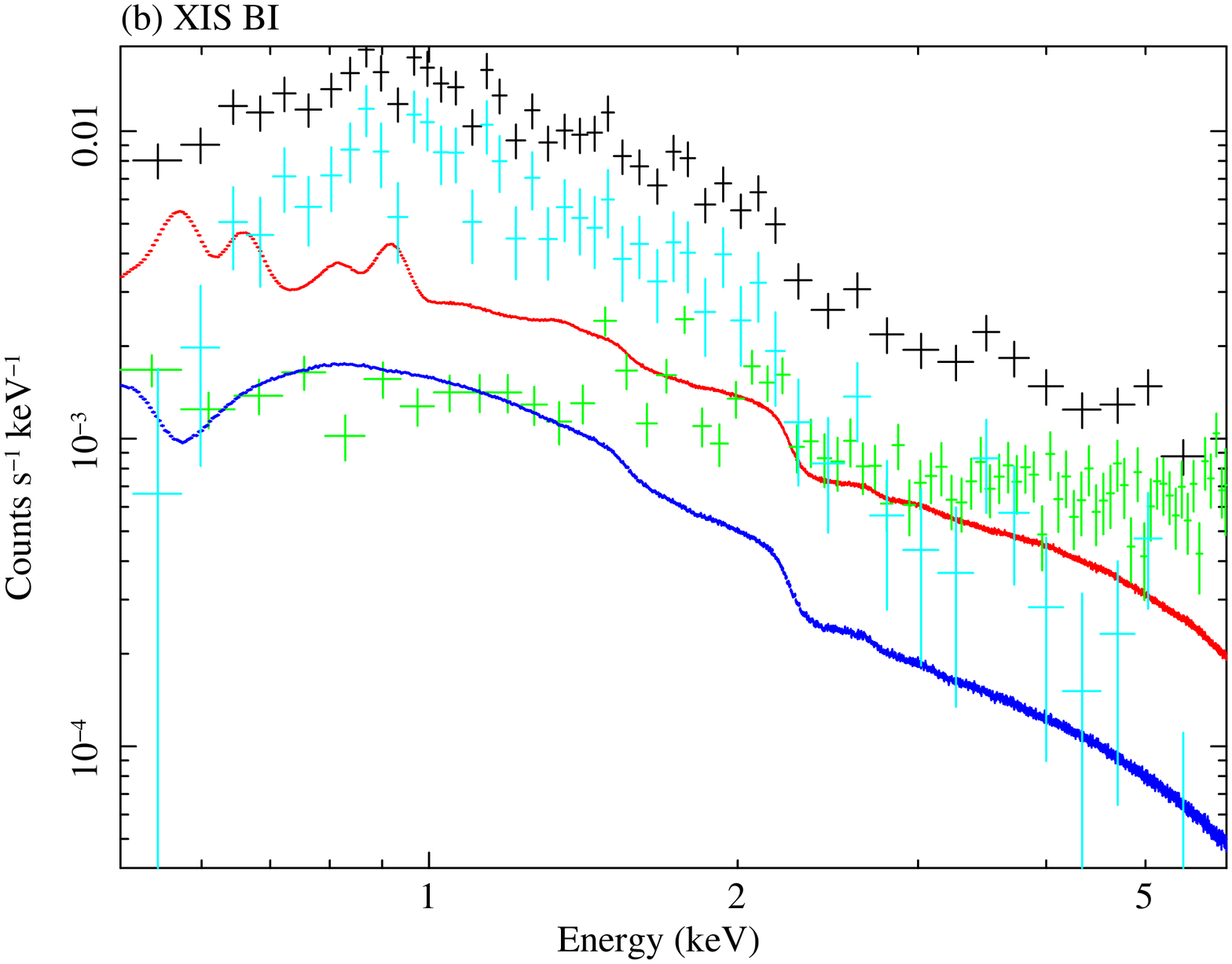}
  \end{center}
  \caption{The raw (black) and the background-subtracted (cyan) spectra of the EE obtained with XIS FI (panel a) and BI (panel b). The long-accumulated non X-ray background and the faked diffuse X-ray background are plotted in green and red respectively. The blue line represents the simulated contamination from the six point sources. }
  \label{fig:bgdsummary}
\end{figure}

\begin{table}
\caption{The $0.5-6$~keV count rates of individual spectral components.}\label{tab:bgdsummary}
\begin{center}
\begin{tabular}{ccc}
\hline
Component & \multicolumn{2}{c}{Rate~($10^{-3}~\countss$)}\\
& XIS FI & XIS BI\\ \hline
Raw & 28.3 & 22.7\\
NXB & 4.4 & 5.3\\
DXB\footnotemark[$*$] & 8.3 & 5.9\\
PS\footnotemark[$\dagger$] & 2.9 & 2.3\\
BGD\footnotemark[$\ddagger$] & 15.6 & 13.5\\
EE\footnotemark[$\S$] & 12.7 (45\%) & 9.2 (40\%)\\
\hline
\multicolumn{3}{@{}l@{}}{\hbox to 0pt{\parbox{60mm}{\footnotesize
\vspace{0.2cm}
\footnotemark[$*$]Diffuse X-ray background.\\
\footnotemark[$\dagger$]Six point sources.\\
\footnotemark[$\ddagger$]Sum of the NXB, diffuse X-ray background, and six point sources.\\
\footnotemark[$\S$]Derived from Raw$-$BGD. Ratios to the Raw count rates are also shown.
}\hss}}
\end{tabular}
\end{center}
\end{table}

\subsection{Model fitting to the Extended Emission Spectrum}
Figure \ref{fig:bgdsummary} shows the raw EE spectra, in comparison with the background components estimated so far.
Table \ref{tab:bgdsummary} summerizes the estimated $0.5-6$~keV count rate of each background component. 
As a whole, the background amounts to about $50\%$ of the raw counts in each detector.
In figure \ref{fig:bgdsummary}, cyan data points indicate the EE spectra obtained after subtracting the three background components. Below, we fit them with several  models which give different physical interpretations.
An ARF for the EE was calculated assuming a uniform circular emitting region with a radius of $\timeform{50''}$ based on the Chandra ACIS imaging result \citep{oka05,oka07}. The FI and BI spectra were fitted simultaneously, with the overall model normalization fixed to be the same between them.

\begin{figure}
  \begin{center}
    \FigureFile(80mm,80mm){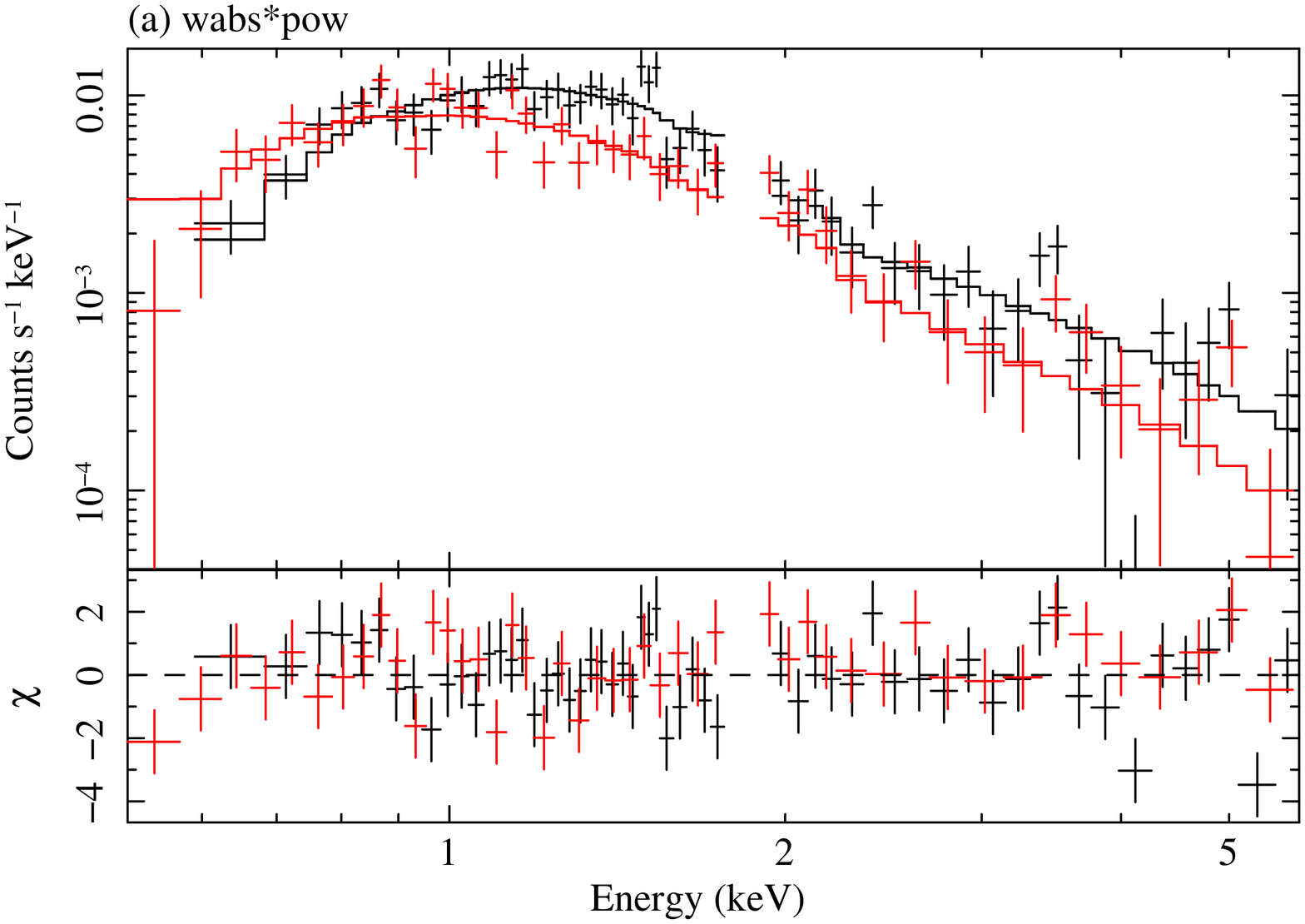}
    \FigureFile(80mm,80mm){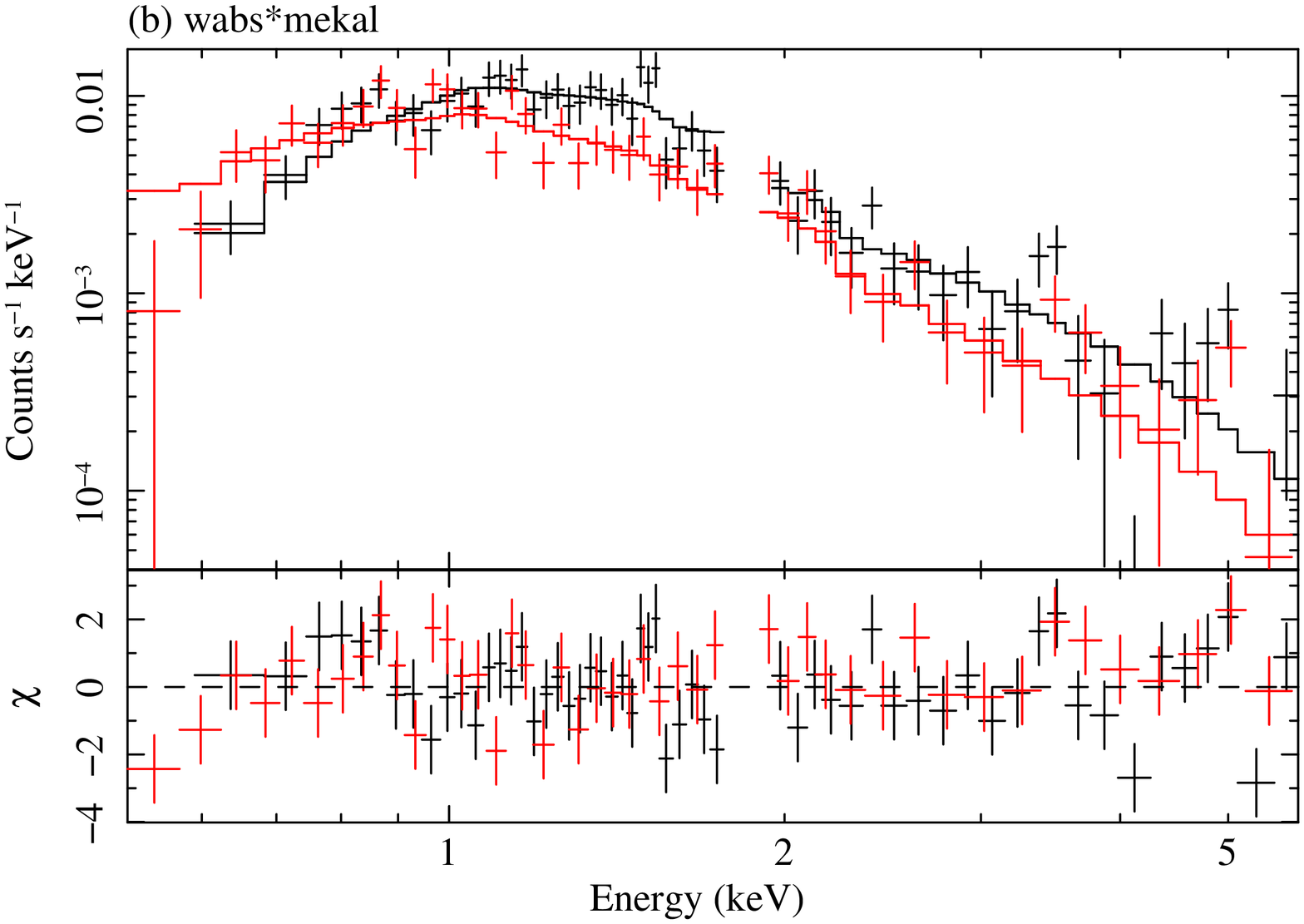}
  \end{center}
  \caption{Spectral fitting to the XIS FI (black) and BI (red) spectra of the EE, with (a) a power-law and (b) a single-temperature thermal models.}
  \label{fig:powmekal}
\end{figure}

First, we fitted the spectra with a single power-law model and a single temperature optically-thin thermal model (\verb|apec| in \verb|XSPEC|; \cite{smi01}), each subjected to the interstellar absorption (\verb|wabs|) as \citet{oka07} did. The fitting results are shown in figure \ref{fig:powmekal} and listed in table \ref{tab:powmekal}.
However, neither the power-law nor optically-thin thermal model reproduced the spectra well, with $\chi^{2}_\nu=1.31~(\nu=100)$ and $1.33~(\nu=99)$ respectively.
The obtained photon index ($\Gamma=2.9\pm0.2$) or the plasma temperature ($kT=1.7\pm0.3$~keV) implies a considerably softer spectral shape than the previous report (\cite{oka05}; $\Gamma=2.1\pm0.3$ or $kT=3.7\pm^{2.7}_{1.3}$~keV).
In section \ref{sec:discussion}, we discuss this discrepancy. 

In figure \ref{fig:powmekal}, we notice some spectral structures around 0.85~keV, 1.5~keV, and 5~keV that cannot be explained by the employed models. Suspecting that these structures originate from redshifted atomic emission lines, we next fitted the spectra with a thermal model that has a free redshift $z$. The fit was improved significantly to $\chi^{2}_\nu=1.10~(\nu=98)$, and yielded the metal abundance and redshift of $0.38^{+0.25}_{-0.13}$ times solar and $z=0.34\pm0.02$, respectively. Especially the spectral features at $\sim0.85$~keV, $\sim1.5$~keV and $\sim5$~keV have been reproduced successfully by redshifted Fe-L, Si-K and Fe-K lines, respectively.
Incorporating $z$ thus determined, the observed flux can be converted to the intrinsic luminosity of $L_{0.5-6~{\rm keV}}=5.5\times10^{43}~\ergs$, $L_{2-10~{\rm keV}}=2.8\times10^{43}~\ergs$, and $L_{0.1-200~{\rm keV}}=1.0\times10^{44}~\ergs$ in the $0.5-6~{\rm keV}$, $2-10~{\rm keV}$, and $0.1-200~{\rm keV}$ band respectively.

Since the XIS background spectrum contains K$\alpha$ emission line from aluminum used in, for example, the XIS housing and substrate of the CCD, the line feature at 1.5~keV could be due to residual Al-K lines caused by a wrong NXB subtraction. To examine this possibility, we also tried spectral fittings with NXB spectra rescaled by $5-10\%$. However, the feature at $\sim1.5$~keV can be seen even after subtracting an NXB spectrum that is rescaled up by $+10$\%. Since the systematic error (or reproducibility) of the NXB estimation is reported to be ~5\% \citep{taw08}, we consider that the structure to be real rather than instrumental. For reference, the fit results remain unchanged within the errors even if we ignore the $1.4-1.6$~keV range in the fitting.

\begin{figure}
  \begin{center}
    \FigureFile(80mm,80mm){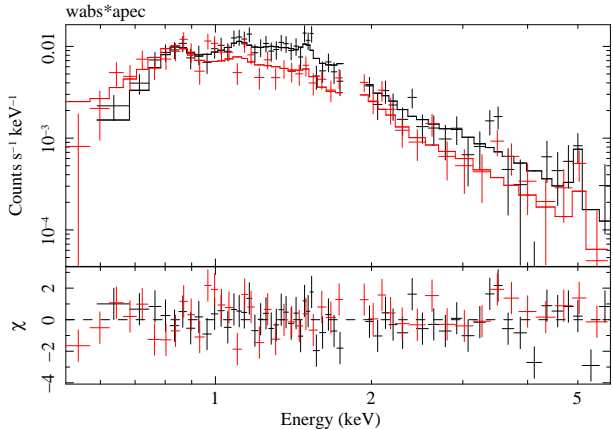}
  \end{center}
  \caption{The same EE spectra as presented in figure \ref{fig:powmekal}, fitted with a redshifted thermal emission model.}
  \label{fig:apec}
\end{figure}

In figure \ref{fig:apec}, we notice fitting residuals both in the XIS FI and BI spectra at ~3.5~keV. However, they have no corresponding background features (figure \ref{fig:bgdsummary}) or redshifted major atomic lines. No such features are present in the Chandra spectrum, either \citep{oka07}. They are hence considered as statistical fluctuations.

\begin{table*}
\caption{The best fit parameters of the EE spectra.}\label{tab:powmekal}
\begin{center}
\begin{tabular}{cccccccc}
\hline
Model & $N_{\rm H}$\footnotemark[$*$] & $\Gamma$ & $kT$\footnotemark[$\dagger$] & $Z$\footnotemark[$\ddagger$] & $z$\footnotemark[$\S$] & $\Sigma$\footnotemark[$\|$] & $\chi^2_\nu~(\nu)$ \\
\hline
Power law & $20^{+6}_{-5}$ & $2.9\pm0.2$ & $-$ & $-$ & $-$ & $7.5^{+1.2}_{-0.9}$ & 1.31 (100)\\
Theraml & $3.3^{+4.4}_{-3.3}$ & $-$ & $1.7\pm0.3$ & $0.02^{+0.06}_{-0.02}$ & $-$ & $7.5^{+1.6}_{-1.3}$ & 1.33 (99) \\
Redshifted thermal& $6.9^{+5.7}_{-4.8}$ & $-$ & $2.2^{+0.2}_{-0.3}$ & $0.38^{+0.25}_{-0.13}$ & $0.34\pm0.02$ & $7.6^{+1.3}_{-1.2}$ & 1.10~(98)\\
\hline
\multicolumn{8}{@{}l@{}}{\hbox to 0pt{\parbox{145mm}{\footnotesize
\vspace{0.2cm}
\footnotemark[$*$]Line-of-sight hydrogen column density in units of $10^{20}~{\rm cm}^{-2}$.\\
\footnotemark[$\dagger$]Thermal plasma temperature in units of keV.\\
\footnotemark[$\ddagger$]Abundance in terms of the solar value.\\
\footnotemark[$\S$]Redshift.\\
\footnotemark[$\|$]The $0.5-6$~keV band model surface brightness in units of $10^{-7}~\ergcmssr$.\\
}\hss}}
\end{tabular}
\end{center}
\end{table*}

The above results strongly suggest that the EE is an extragalactic object, rather than a source associated with 47 Tuc. Further considering the extended nature and the thermal spectrum, it is most likely a background cluster of galaxies at $z=0.34$. In the following section, we examine the galaxy cluster interpretation of the EE based on the Suzaku results.

\section{Discussion}\label{sec:discussion}

\subsection{$kT-L_{\rm{X}}$ relation and the counterpart in other wavelength}
\begin{figure}
  \begin{center}
    \FigureFile(80mm,80mm){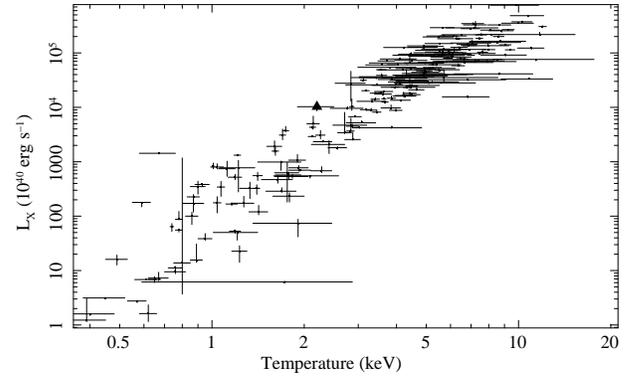}
  \end{center}
  \caption{A $kT-L_{\rm{X}}$ relation of clusters of galaxies. Crosses represent temperatures and bolometric luminosities of individual galaxy clusters determined by ASCA observations (data taken from \cite{fuk04}). Luminosities were obtained by integrating fluxes over the $0.1-200$~keV band. Filled triangle shows the EE of 47 Tuc.}
  \label{fig:ktlx}
\end{figure}

As shown so far,  the spectra of the EE are well described by thermal plasma emission with a rest-frame temperature of $kT=2.2$~keV and a redshift of $z=0.34$. Furthermore, as plotted in figure \ref{fig:ktlx}, its luminosity and temperature are consistent with the known temperature-luminosity relation ($kT-L_{\rm{X}}$ relation) of cluster of galaxies. Therefore, the EE is most naturally interpreted as a background cluster of galaxies at a moderate redshift.

We find no counterpart in the optical (Digital Sky Survey; e.g. \cite{mcl00}) or near infrared (Two Micron All Sky Survey; \cite{skr06}) surveys. Using Chandra deep survey data, \citet{bos02} however reported more than 20 candidates of clusters of galaxies that have no optical counterpart.
The present background galaxy cluster is perhaps a member of those clusters. A deeper optical imagery will reveal the expected galaxy clustering.

\subsection{Comparison with previous reports on the EE}
The galaxy cluster interpretation has been examined as an origin of the EE in previous papers as well. Using the $\log N-\log S$ relation of galaxy clusters, \citet{kg95} and \citet{oka07} estimated probabilities of a chance coincidence of the EE and a background galaxy cluster emission to be less than 0.5\% and 0.6\%, respectively. Based on such low probabilities, these authors argued that the EE cannot be a background galaxy cluster.

In addition to the above probability estimation, \citet{oka07} used the following argument to rule out the background cluster interpretation. First, they determined the EE temperature as $kT=3.7~$keV from the Chandra ACIS spectrum. They hence assigned a luminosity of $L_{\rm{X}}=1.1\times10^{44}~\ergs$ to this putative cluster, using the $kT-L_{\rm{X}}$ relation of clusters (e.g. \cite{ike02,fuk04}). Comparing this $L_{\rm{X}}$ with the measured flux, the source redshift was estimated as $z>0.5$, and hence the observed angular core radius of the EE, $r_{\rm{c}}\sim\timeform{0.6'}$, was converted to a physical size of $r_{\rm{c}}>360~$kpc. Finally, they concluded this $r_{\rm{c}}$ to be too large for a cluster.

In the present study, the use of the Suzaku XIS has enabled us to achieve two major improvements (or revisions) over \citet{oka07}. One is that we clearly detected redshifted emission lines, which indicate $z=0.34\pm0.02$; the Chandra data gave no constraint on $z$. The other is that we measured a significantly lower temperature, $kT=1.7~$keV if assuming $z=0$, or $kT=2.2~$keV at the rest frame if adopting $z=0.34$; the latter now satisfies the $kT-L_{\rm{X}}$ relation of clusters of galaxies (figure \ref{fig:ktlx}). In addition, using the redshift, the physical core radius is now calculated to be $\sim160$~kpc, which is reasonable for galaxy clusters.

As reviewed so far, the difference of our conclusion from that of \citet{oka07} comes mainly from the discrepant EE temperatures, $kT=1.7\pm0.3~$keV measured with Suzaku (without correction for the redshift) and $kT=3.7^{+2.7}_{-1.3}~$keV with the Chandra ACIS. Possible causes of this difference include an over estimation of the temperature with Chandra, or an under estimation with Suzaku, or both. As the former possibility, the most likely cause is systematic errors in the NXB subtraction. As the latter possibility, we may presume that during the Suzaku observation, some soft sources became brighter than in the Chandra observation.

Although the Suzaku data could thus be under-estimating the EE temperature, significantly higher values of $kT$ would be still consistent with the $kT-L_{\rm{X}}$ relation. Furthermore, the value of $z=0.34$ is not affected, since it is determined by the redshifted atomic emission lines.
We conclude that the close spatial coincidence between the EE and 47 Tuc is accidental, and they are not physically associated with each other.

\ \\
\ \\
T.Y. is financially supported by the Japan Society for the Promotion of Science. This research has made use of data and softwares obtained from the Data Archive and Transmission System at JAXA/ISAS and the High Energy Astrophysics Science Archive Research Center, provided by NASA's Goddard Space Flight Center respectively. We obtained the Chandra data from the Chandra Data Archive, and analyzed them with softwares provided by the Chandra X-ray Center.
The present research is supported in part by the Grant-in-Aid for Scientific Research (S), No. 18104004.

\end{document}